\documentclass[preprint,12pt]{elsarticle}

\usepackage{amssymb}
\usepackage{amsmath}
\usepackage{hyperref}
\usepackage{natbib}
\usepackage{caption}
\usepackage{adjustbox}
\usepackage{multirow}
\usepackage{booktabs}
\usepackage{arydshln}
\usepackage{utfsym}
\usepackage{float}

\journal{Biomedical Signal Processing and Control}

\begin{document}

\begin{frontmatter}



\title{MTS-Net: Dual-Enhanced Positional Multi-Head Self-Attention for 3D CT Diagnosis of May-Thurner Syndrome}


\tnotetext[cor1]{Corresponding author.}
\author[label1]{Yixin Huang}
\author[label2]{Yiqi Jin}
\author[label3]{Ke Tao\corref{cor1}}\ead{20234155006@stu.suda.edu.cn}
\author[label3]{Kaijian Xia}
\author[label3]{Jianfeng Gu}
\author[label4]{Lei Yu}
\author[label1]{Haojie Li}
\author[label1]{Lan Du}
\author[label1]{Cunjian Chen\corref{cor1}}\ead{cunjian.chen@monash.edu}

\affiliation[label1]{organization={Department of Data Science and AI, Monash University}, 
            addressline={}, 
            city={Melbourne},
            postcode={3800}, 
            state={Victoria},
            country={Australia}}
            
\affiliation[label2]{organization={Department of Vascular Surgery, The Affiliated Suzhou Hospital of Nanjing Medical University}, 
            addressline={}, 
            city={Suzhou},
            postcode={215008}, 
            state={},
            country={China}}
            
\affiliation[label3]{organization={Department of Vascular Surgery, Changshu Hospital Affiliated to Soochow University}, 
            addressline={}, 
            city={Changshu},
            postcode={215500}, 
            state={},
            country={China}}
            
\affiliation[label4]{organization={Central Laboratory, Guiyang Maternal and Child Health Hospital}, 
            addressline={}, 
            city={Guiyang},
            postcode={550003}, 
            state={},
            country={China}}

\begin{abstract} 
May-Thurner Syndrome (MTS) is a vascular condition that affects over 20\% of the population and significantly increases the risk of iliofemoral deep venous thrombosis. Accurate and early diagnosis of MTS using computed tomography (CT) remains a clinical challenge due to the subtle anatomical compression and variability across patients. In this paper, we propose MTS-Net, an end-to-end 3D deep learning framework designed to capture spatial-temporal patterns from CT volumes for reliable MTS diagnosis. MTS-Net builds upon 3D ResNet-18 by embedding a novel dual-enhanced positional
multi-head self-attention (DEP-MHSA) module into the Transformer encoder of the
network's final stages. The proposed DEP-MHSA employs multi-scale convolution and
integrates positional embeddings into both attention weights and residual paths, enhancing spatial context preservation, which is crucial for identifying venous
compression. To validate our approach, we curate the first publicly available dataset for
MTS, MTS-CT, containing over 747 gender-balanced subjects with standard and
enhanced CT scans. Experimental results demonstrate that MTS-Net achieves average 0.79
accuracy, 0.84 AUC, and 0.78 F1-score, outperforming baseline models including 3D
ResNet, DenseNet-BC, and BabyNet. Our work not only introduces a new diagnostic
architecture for MTS but also provides a high-quality benchmark dataset to facilitate
future research in automated vascular syndrome detection. 
 We make our code and dataset publicly available at: 
 \underline{\url{https://github.com/Nutingnon/MTS_dep_mhsa}}. 
 
\end{abstract} 


    
    
    
    


\end{frontmatter}



\section{Introduction} 
\label{sec：introduction}
{M}{ay-Thurner} Syndrome (MTS) is a vascular disorder in which the left common iliac vein is abnormally compressed between the right common iliac artery and the underlying lumbar vertebra, impeding venous outflow from the lower limb~\cite{mts_knlg_wolpert2002magnetic}. Type I MTS refers to an anatomical variant in which the right common iliac artery overlies and compresses the left common iliac vein against the lumbar vertebrae. It exists in more than 20\% of the population, potentially leading to iliofemoral deep venous thrombosis (DVT) in patients, with a 2\% to 3\% chance of occurrence~\cite{medical_knowledge_kalu2013may, peters2012may_medical_knowledge}. MTS diagnosis predominantly relies on medical professionals' expertise through advanced imaging techniques, including venography and intravascular ultrasound (IVUS), ultrasound (US), computed tomographic (CT) or magnetic resonance imaging (MRI)~\cite{mts_knlg_wolpert2002magnetic}. Each imaging technology has its own strengths and limitations. The gold standard for the MTS diagnosis is IVUS~\cite{medical_knowledge_Ultrasound_point}. However, this technique is invasive and costly, and the contrast agents may lead to allergic reactions, phlebitis or post-injection DVT .  On the other hand, MRI has better contrast resolution and both MRI and US have no radiation hazards~\cite{knuttinen2017may}. However, from patients' perspective, MRI always leads to much higher cost in price and time during examination than other methods. While ultrasound is considered the least expensive imaging method, it is highly sensitive and adept at assessing proximal lower extremity DVT. Therefore, ultrasound demonstrates limited sensitivity above the inguinal plane, making it unfeasible to evaluate the compressibility of the iliac vein~\cite{knuttinen2017may, oguzkurt2007ultrasonographic}. However, despite the radiation exposure associated with CT imaging, it demonstrates high sensitivity and specificity for the entire region. Additionally, it enables the reconstruction of multiple plane images with excellent details \cite{knuttinen2017may}. Medical experts often resort to Enhanced-CT examinations, which involve injecting a significant volume of contrast agent into the veins, to achieve clearer differentiation between the vein and adjacent tissues. 

Meanwhile, there is also limited research work on applying deep learning-based methods to MTS diagnosis. A notable study employed the DMRF-CNN~\cite{mts_rcnn} to train on a private ultrasound dataset, consisting of 34 male and 177 female patients. 
However, due to inherent limitations in using ultrasound imaging for MTS diagnosis and restricted access to private datasets in existing studies, there is a significant opportunity to leverage recent advances in deep learning for MTS diagnosis. Given the high prevalence of MTS and the limitations of existing imaging-based diagnostic tools, there is a clear need for computational approaches that can assist clinicians in recognizing subtle venous compression patterns from CT images. With the growing success of deep learning in various vascular and structural imaging tasks, it becomes timely and meaningful to explore dedicated solutions tailored for MTS diagnosis. On the other hand, AI-driven methods in medical image recognition thrive with the high-quality datasets that are continuously being collected and published.
This is evidenced by annual challenges such as the brain tumor radiogenomic classification~\cite{miccai21_bakas2021rsna},  intracranial hemorrhage detection~\cite{rsna-intracranial-hemorrhage-detection-2019}, and cervical spine fracture detection~\cite{rsna-2022-cervical-spine-fracture-detection}. The availability of these rich datasets has significantly advanced deep learning for intelligent diagnosis in specific medical areas. While deep learning-based methods are widely applied in medical diagnosis, there is limited research specifically targeting May-Thurner Syndrome diagnosis, primarily due to the scarcity of public datasets for MTS diagnosis. Furthermore, unlike other vascular imaging tasks, MTS diagnosis relies on detecting subtle, localized compressions in full-resolution CT scans—patterns that general-purpose models often fail to capture. Existing research lacks dedicated solutions for this challenge. To address this gap, we propose a task-specific neural architecture and introduce a benchmark dataset to support further development in this underexplored area.

In this paper, we present MTS-Net, an end-to-end 3D neural network architecture designed to effectively model spatial-temporal relationship for improving MTS diagnosis using 3D CT scans. Specifically, MTS-Net enhances the vanilla 3D ResNet-18 by integrating a novel self-attention module, termed DEP-MHSA, within the Transformer encoder block at the network's final two layers. The DEP-MHSA module is meticulously designed to employ a threefold convolution strategy, enabling it to extract features from multiple scales. It further amplifies the model's capability by incorporating dual-enhanced positional embeddings at critical computation stages. These positional embeddings play a significant role in preserving the spatial context of the CT scans, where the relative positions of anatomical structures can be vital for a correct diagnosis. This design reflects the diagnostic strategy employed by radiologists when visually interpreting anatomical positions and relationships in CT slices, and is purposefully aligned with the objective of detecting fine-grained structural changes unique to MTS cases. This architecture is deliberately crafted to imitate the diagnostic strategies employed by medical experts in analyzing CT scans for MTS, thereby offering greater precision and adaptability. To further validate the effectiveness of the proposed approach, we conduct experiments on a comprehensive and gender-balanced CT dataset encompassing over 700 subjects. To the best of our knowledge, the dataset collected in this study is the first publicly accessible dataset for MTS diagnosis using 3D CT scans. The main contributions are summarized as follows:
\begin{itemize}
    \item We propose a novel framework called MTS-Net, which extends 3D ResNet-18 with a novel residual transformer module for May-Thurner Syndrome diagonis using 3D CT scans. 
    \item We propose a new multi-head self-attention module called DEP-MHSA that effectively integrates dual-enhanced positional embeddings that emulates the clinical process of diagnosing MTS.
    \item We introduce the first publicly accessible MTS dataset, comprising standard and Enhanced-CT scans. This dataset serves as a foundational resource for future research in this domain.
    \item We evaluate the performance of our method comprehensively and demonstrate superior accuracy against existing methods on the proposed dataset.
\end{itemize}
The rest of the paper is organized as follows. Section~\ref{relatd_works} reviews recent work on 3D CNNs and Transformers and their applications in medical image analysis. Section~\ref{method} introduces the proposed MTS-Net method, with emphasis on the DEP-MHSA and its variants. Section~\ref{main_exp} presents the new self-collected MTS-CT dataset, as well as discussing the experiments and results related to the May-Thurner Syndrome diagnosis. Section~\ref{discussion} offers additional experiments to validate the effectiveness of using Enhanced-CT scans. Section~\ref{conclusions} concludes the work.

\section{Related Work}
\label{relatd_works}
\subsection{3D CNNs.} 
3D Convolutional Neural Networks (3D CNNs) have shown remarkable success in a wide range of applications involving three-dimensional data~\cite{r2dplus1d}. These networks extend 2D CNNs by adding an additional dimension, enabling them to process 3D data effectively. The architecture of 3D CNNs provides a strong inductive bias on capturing local feature relationships in three dimensions. This capability is particularly crucial in medical imaging, where tumors or other diseased tissues often appear relatively small compared to the entire scanned regions. For instance, in the field of coronavirus disease diagnosis and prognosis management, researchers have employed 3D CNNs to recognize local anomaly patterns on lung CT scans~\cite{cnn_covid19_example1}, predict the prognosis situation\cite{fibro_lung} and identify brain tumors in MRI scans\cite{deepak2019cnn_for_brain}. Additionally, 3D-based UNet~\cite{unet} have been successfully used in medical image segmentation. SimU-Net \cite{liver_segmentation_unet_2023}, designed for liver lesion segmentation, incorporated 3D CNNs with varying kernel sizes and modified the placement of residual connections for enhanced performance. Moreover, the method introduced in \cite{ipmi2023_brain_deeplearning} employed carefully designed 3D CNNs within an encoder-decoder network to accurately model the shape of the brain connectome. Our work is fundamentally built upon existing 3D CNNs, enhanced with a novel spatial-temporal attention module. 

\subsection{Transformers.} 
Transformers\cite{attention_is_all} and the associated self-attention mechanisms\cite{shaw2018selfAttnRelativePosition} have revolutionized the field of deep learning in recent years. Initially developed for natural language processing, Transformers-based technologies~\cite{xu2023levit, vit3d, vivit, plotka2022babynet} have demonstrated remarkable capabilities in medical imaging analysis across various tasks such as segmentation, classification, regression, as well as different modalities including CT, MRI, and X-ray~\cite{Transformer_medical_image_survey2023, medical_image_review_on_attention}. The self-attention mechanism, an integral component of the Transformer architecture, excels at modelling complex relationships among tokens in sequences. This is particularly beneficial in medical imaging analysis, where accurately modeling the relationships between anatomical structures is essential for precise diagnosis. For instance, Meta-ViT was introduced for the diagnosis of Parkinsonism \cite{MetaViT_IPMI2023}, which adopts a vision transformer network with metabolism-aware blocks. These blocks incorporated a standard vision transformer layer enhanced by a novel inter-patch voxel-wise self-attention mechanism. Nevertheless, Transformers lack the inductive bias in modeling local structures. Therefore, some works directly incorporate a self-attention mechanism into models such as UNet \cite{unet} or ResNet to better explore such intrinsic inductive bias, as well as capturing long-range interactions. In parallel, recent efforts have explored Transformer variants in vascular imaging tasks. For instance, IRON~\cite{pan2024iterative} introduced a convolution-assisted Transformer for limited-angle cardiac CT reconstruction, addressing image synthesis challenges via residual optimization. Meanwhile, MALAR~\cite{zhang2022multiple} applied adversarial learning and cross-domain constraints to enhance vascular visibility in ultra-low-dose angiography. However, both methods are fundamentally designed for image reconstruction and enhancement, rather than for diagnosing anatomical compression syndromes such as MTS. In this work, we take advantage of both Transformers and CNN structures for 3D CT Diagnosis of May-Thurner syndrome.

\section{Method} 
\label{method}
In this section, we present our method MTS-Net, which introduces dual-enhanced positional multi-head self-attention in the residual transformer module, as shown in Fig.~\ref{fig_network1}. We utilize 3D ResNet-18 consisting of a 2D spatial convolution followed by a 1D temporal convolution as the backbone~\cite{r2dplus1d}. Further, the final two layers are adapted to incorporate the proposed dual-enhanced positional multi-head self-attention block, as illustrated in Fig.~\ref{dep_mhsa_fig}. The proposed network draws macro-structural inspiration from BabyNet~\cite{plotka2022babynet}, which utilizes fetal ultrasound data to predict birth weight by employing a 3D ResNet-based network with the last layer substituted by its proposed residual transformer encoder block. At a micro level, the proposed module in this study draws inspiration from the diagnostic reasoning employed by medical experts in evaluating May-Thurner Syndrome through CT images. Specifically, medical experts first identify key frames from a complete set of CT scan images. They then analyze the spatial-temporal relationship between veins and bones within these key frames to arrive at a conclusive diagnosis. Therefore, the DEP-MHSA block mirrors their approach by placing distinct emphasis on the functions generating the Query, Key, and Value matrices in self-attention mechanism. This enhancement improves the model's diagnostic precision and aligns well with expert clinical procedures.

\subsection{Model Architecture} \label{model_architecture}
Our proposed MTS-Net reuses the convolutional stem from the 3D ResNet-18~\cite{r2dplus1d} by decomposing 3D convolution into 2D spatial and 1D temporal convolutions, referred to as the (2+1)D convolution. This stem is structured as two sequential operations of convolution, batch normalization, and ReLU (Conv-BN-ReLU) in the (2+1)D  configuration, designed for the initial extraction of spatial-temporal features from CT scans.

\begin{figure*}[ht]
\centering
\includegraphics[width=\textwidth]{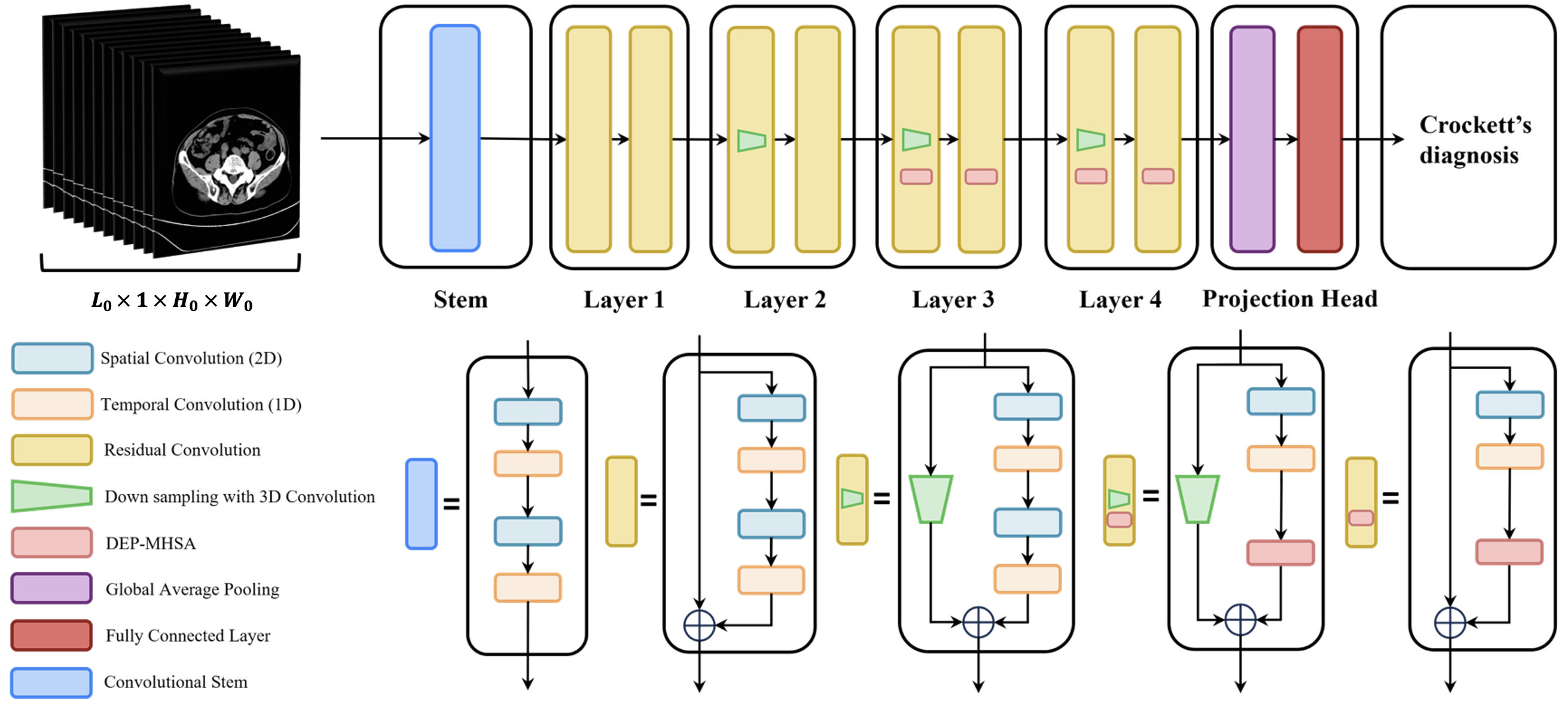}
\caption{The overview illustrates our approach for diagnosing May-Thurner Syndrome from the input CT scan videos. This methodology is anchored in the modified ResNet-18 (2+1)D architecture. Uniquely, we have transformed the standard Spatial-Temporal convolution in the final two layers into our proposed Multi-Head Self-Attention module. The first row of the figure shows the overall structure of the network from input to output, and the five diagrams to the right of the second row show the more detailed composition of each block in the overall structure diagram corresponding to the first row. This designed adaption is inspired by the logic behind the diagnosing process of medical experts, thereby significantly improving performance in identifying May-Thurner syndrome.}
\label{fig_network1}
\end{figure*}

Let $x_{0}$ denote the input CT scan clip, represented as a tensor in the space $\mathbb{R}^{L_{0} \times C_{0} \times H_{0} \times W_{0}}$. Here, the dimensions of the tensor correspond to specific characteristics of the CT scan. $L_{0}$ represents the length of the clip in terms of the number of frames, capturing the temporal aspect of the scan. $C_{0}$ denotes the channel of the clip, with $C_{0}=1$ reflecting the grayscale nature of CT images. $H_{0}$ and $W_{0}$ denote the height and width of each frame, respectively, providing the spatial dimensions of the scan. This multi-dimensional representation is widely used when processing CT scans in medical imaging. Given the input CT clip $x_{0}$, the computational sequence at the $i$-th layer is represented as $\mathcal{F}^{(i)}(\cdot;\theta^{(i)})$, where $\mathcal{F}^{(i)}$ denotes the functional operations and $\theta^{(i)}$ signifies the associated learnable parameters. Specifically, the resulting output from the convolutional stem layer can be expressed as:
\begin{equation}
\label{formula_stem}
x_{1} = \mathcal{F}^{(0)}_{T}(\mathcal{F}^{(0)}_{S}(x_{0} ; \theta_{S}) ; \theta_{T}).
\end{equation}
Here, $\mathcal{F}_{S}$ refers to the spatial computation sequence: 
\begin{equation}
\label{f_spatial}
\mathcal{F}_{S}(x;\theta_{S}) = \sigma(BN(Conv(x; 1,k,k))),  \\
\end{equation}
where, $Conv$ denotes the 2D spatial convolution, $BN$ denotes batch normalization, $\sigma$ represents non-linear activation function ReLU and $k$ is the kernel size. On the other hand,  $\mathcal{F}_{T}$ indicates temporal computation sequence: 
\begin{equation}
\label{f_temporal}
\mathcal{F}_{T}(x;\theta_{T}) = \sigma(BN(Conv(x; k,1,1))),
\end{equation}

where $Conv$ denotes the 1D temporal convolution instead.
Such a combination, as indicated by Eq.~\eqref{formula_stem}, effectively forms what we refer to as the spatial-temporal structure of the convolutional stem. Then, we apply a series of layers, each layer is composed of two residual blocks. For first two layers, each residual block generally contains two spatial-temporal operation with residual connection:

\begin{equation}
\label{formula_layer1}
x_{i} = \mathcal{F}^{(i)}(x_{i-1}; \theta^{(i)}) + x_{i-1},
\end{equation}
where $x_{i}$ denotes the intermediate output from the $(i-1)$th layer $\mathcal{F}^{(i-1)}$ after the convolutional stem layer, i.e., $i>1$. 
Note that the output from Layer 1, depicted in Fig.~\ref{fig_network1}, retains the same height and width as the input clips. Specifically, Layer 1 is comprised of two residual spatial-temporal convolution blocks, similar to the one used in convolutional stem but containing a residual connection. Starting from the Layer 2, the height and width of the output undergo downsampling by a factor of 2 at each subsequent layer. Consequently, the dimensions of the resulting feature map from Layer 4 are reduced to $H_{0}/8$ for height and $W_{0}/8$ for width, respectively. Parallel to this spatial downsampling, the temporal dimension, denoted as $L_{0}$ from the input, is also progressively reduced, culminating in a downsampling to $L_{0}/4$ by the last residual layer. The projection head is composed of a global average pooling (GAP) operation and a fully-connected layer which maps the intermediate output from Layer 4 to final predicted output. 

\subsection{Dual-Enhanced Positional Multi-Head Self-Attention}
\begin{figure*}[ht]
\centering
\includegraphics[width=0.7\textwidth]{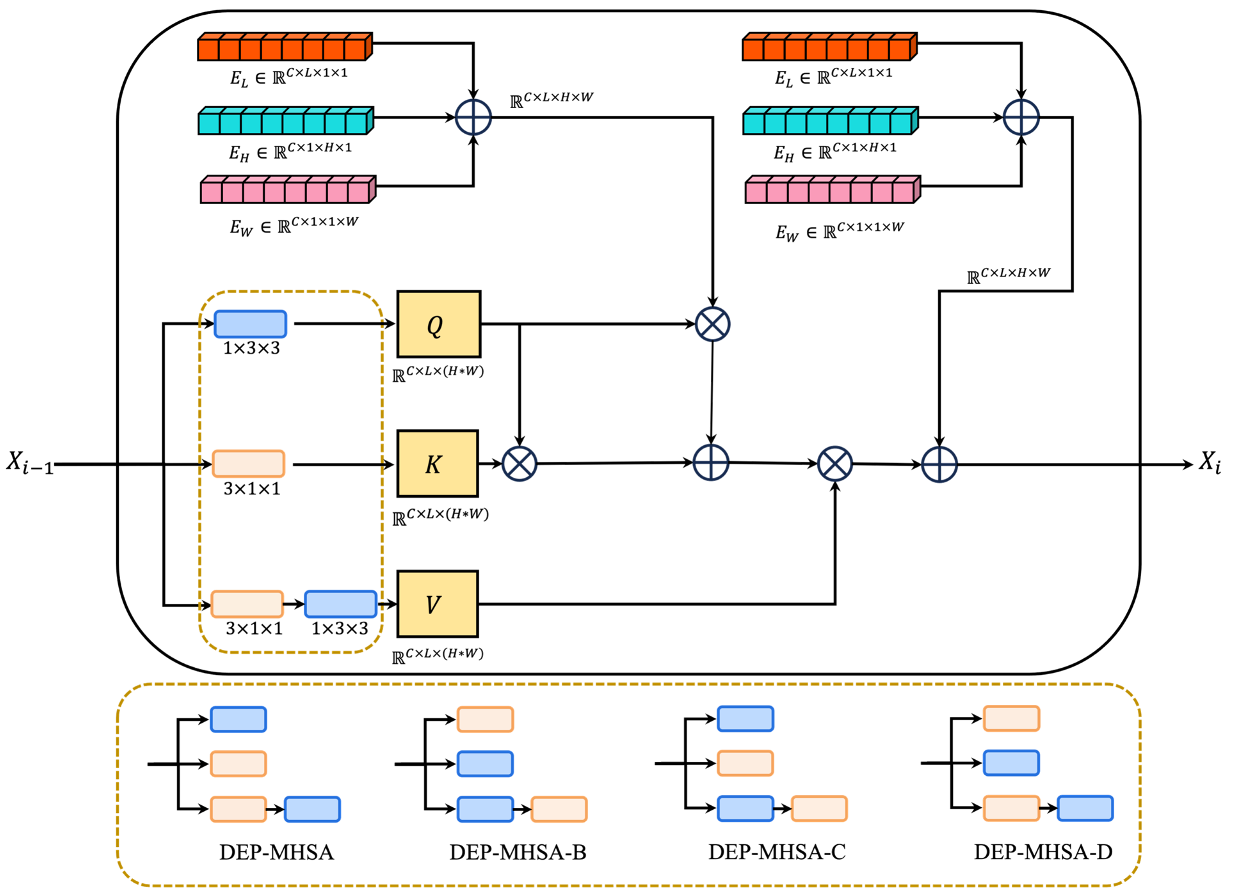}
\caption{A detailed schematic representation of the dual-enhanced positional multi-head self-attention (DEP-MHSA) module. The highlighted yellow dotted region illustrates the critical component where input $x$ is mapped to the Query ($Q$), Key ($K$), and Value ($V$) matrices for the self-attention computation. It can have 4 variants as listed. The symbols $E_{L}$, $E_{H}$ and $E_{W}$ denote the relative position embeddings corresponding to frames, height, and width, respectively. This first set of relative position embeddings participates to the calculation of $Q K^{T}$, and the second set of relative position embeddings behave as residual connection.}
\label{dep_mhsa_fig}
\end{figure*}

The multi-head self-attention mechanism often requires prohibitive computational resources for long input sequences. In our experiments, the input size for layer 1 and layer 2 are $12\times128\times128$ and $12\times64\times64$ (representing frames, height, and width, respectively), which become excessively large when flattened for self-attention processing. Consequently, we put our novel self-attention module, DEP-MHSA, in layers 3 and 4, rather than in the initial layers. Traditionally, most studies that incorporate self-attention mechanism for visual tasks have employed $1\times1\times1$ convolution to map the input to the Query, Key and Value matrices \cite{plotka2022babynet, ruan2022mmdiffusion}. However, drawing the inspiration from the clinic process in diagnosing May-Thurner Syndrome, the proposed self-attention block, illustrated in Fig.~\ref{dep_mhsa_fig}, adopts different configurations for generating the Query, Key, and Value matrices, where $Q=W_{Q}^{(i)} \cdot \varphi(z)$, $K=W_{K}^{(i)} \cdot \tau(z)$ and  $V=W_{v}^{(i)} \cdot \rho(z)$. Here, $W_{Q}$, $W_{K}$ and $W_{V}$ represent distinct weight matrices that transform the input into Query ($Q$), Key ($K$) and Value ($V$). $\varphi(z)$, $\tau(z)$ and $\rho(z)$ are flattened intermediate representation in the space $\mathbb{R}^{l \times c \times (h*w)}$.

Besides, we significantly amplify the role of positional embeddings through both the computation of attention weights and residual connection to the block's output. Specifically, the strategically refined configuration of multi-head self-attention is described as follows:   
\begin{itemize}
  \item The $\varphi(\cdot)$ is a 3D convolution operation with $1\times3\times3$ kernel size for frames, height, and width respectively. The $\tau(\cdot)$ is a $3\times1\times1$ convolution. For the $\rho(\cdot)$, we use (1+2)D settings that a $3\times1\times1$ convolution followed by a $1\times3\times3$ convolution. This design emulates the medical diagnosis process of experts, as they quickly select key frames by going through the CT scan video (Query). The experts then take a closer look at these key frames (Key) and speculate on the position and relationship of human tissue in the three-dimensional space based on the order and content of each frame (Value). 
  
  \item To fully utilize the relative position information, we draw lessons from the popular self-attention practices \cite{vit3d, attention_is_all, shaw2018selfAttnRelativePosition, zhou2021deepvit} that add relative position embeddings to the output. Besides, we also consider fusing the relative positional embeddings into the attention weight calculation process \cite{transformer_bottleneck, plotka2022babynet}. 
\end{itemize}

\noindent\textbf{Relative Position Encoding.} The relative position is encoded across three dimensions: height, width, and time frames. Let $E_H \in \mathbb{R}^{C \times 1 \times H \times 1}$, $E_W \in \mathbb{R}^{C \times 1 \times 1 \times W}$, and $E_F \in \mathbb{R}^{C \times L \times 1 \times 1}$ represent the height, width, and frame position encodings, respectively. The overall relative position encoding is finally obtained as:
\begin{equation}
E = E_{H} + E_{W} + E_{F},
\end{equation}
where each \(E_\ast\) is a learnable tensor that varies only along its dedicated axis. Both the frame‑wise and spatial relative‑position matrices are initialised with $\mathcal{N}(0,0.02)$ and updated end‑to‑end during training.

\noindent\textbf{Transformation for Multi-Head Attention.} The encoded tensor $E$ is reshaped to facilitate multi-head attention, forming a new tensor $M$, which is in shape of $(F, H\times W, \frac{C}{n_{head}})$ where $F$, $H$, $W$, and $n_{head}$ denote the number of frames, height, width, and the number of attention heads, respectively.
\noindent\textbf{Multi-Head Self-Attention.} The self-attention mechanism for each head is computed as follows. Let $Q_i$, $K_j$, and $V_j$ denote the query, key, and value vectors. The attention scores $e_{ij}$ between each query $Q_i$ and key $K_j$, modified by the relative position matrix $M_j$, are calculated by:
\begin{equation}
  e_{ij} = \frac{Q_{i} \cdot (K_{j}^{T}+M_{j})}{\sqrt{d}},
\end{equation}
  where $d$ is the scaling factor, typically the dimensionality of the key vectors. The attention weights $a_{ij}$ are then obtained using the softmax function:
\begin{equation}
  a_{ij}  = \frac{exp(e_{ij})}{\sum_{j=1}^{N}exp(e_{ij})}.
\end{equation}
  Subsequently, the output vector $Y_i$ for each query is derived by aggregating over all values $V_j$, weighted by the computed attention weights:
\begin{equation}
  Y_{i} = \sum_{j=1}^{N}a_{ij}V_{j},
\end{equation}
where $N$ is the total number of keys and values.

\noindent\textbf{Output Reshaping and Final Adjustment.} The resulting attention outputs are reshaped back to the original dimensions and combined with the original position encodings: $output = Y+E$, where $Y \in \mathbb{R}^{C \times F \times H \times W}$ represents the output of the attention mechanism, and the addition operation integrates the learned attention with the original relative position encodings.

\subsection{Model Variants}
To ascertain whether the effectiveness of DEP-MHSA stems from its medically-inspired logic rather than from an increased parameter size, we evaluate its performance across permutations with similar logic but different sequence in generating matrices Query, Key, and Value, as depicted in Fig.~\ref{dep_mhsa_fig}. We rigorously tested variants named DEP-MHSA-B, DEP-MHSA-C, and DEP-MHSA-D. Specifically, DEP-MHSA-B alternates the convolution settings for generating Query and Key, and reverses the convolution sequence in Value generation. DEP-MHSA-C solely reverses the convolution sequence in Value generation, while DEP-MHSA-D exclusively swaps the convolution settings between Query and Key.

\section{Experiments}
\label{main_exp}
\subsection{Dataset and Evaluation Metrics}
\begin{figure}[ht!]
\centering
\includegraphics[width=0.6\columnwidth]{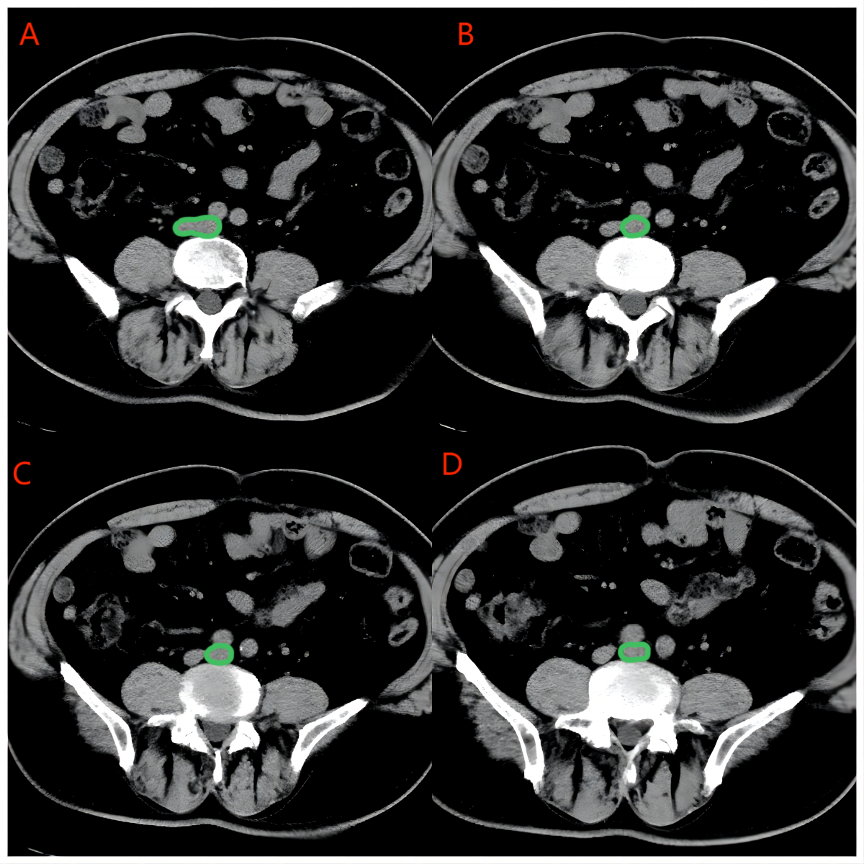}
\caption{Pic. A: Green mark is the bifurcation of the inferior vena cava, which is divided into left and right common iliac vein; Pic. B-D: The green mark is the left common iliac vein, which is visible between the surrounding right common iliac artery and the fifth lumbar spine, with no obvious compression stenosis.}
\label{cockett_light}
\end{figure}

\begin{figure}[ht!]
\centering
\includegraphics[width=0.6\columnwidth]{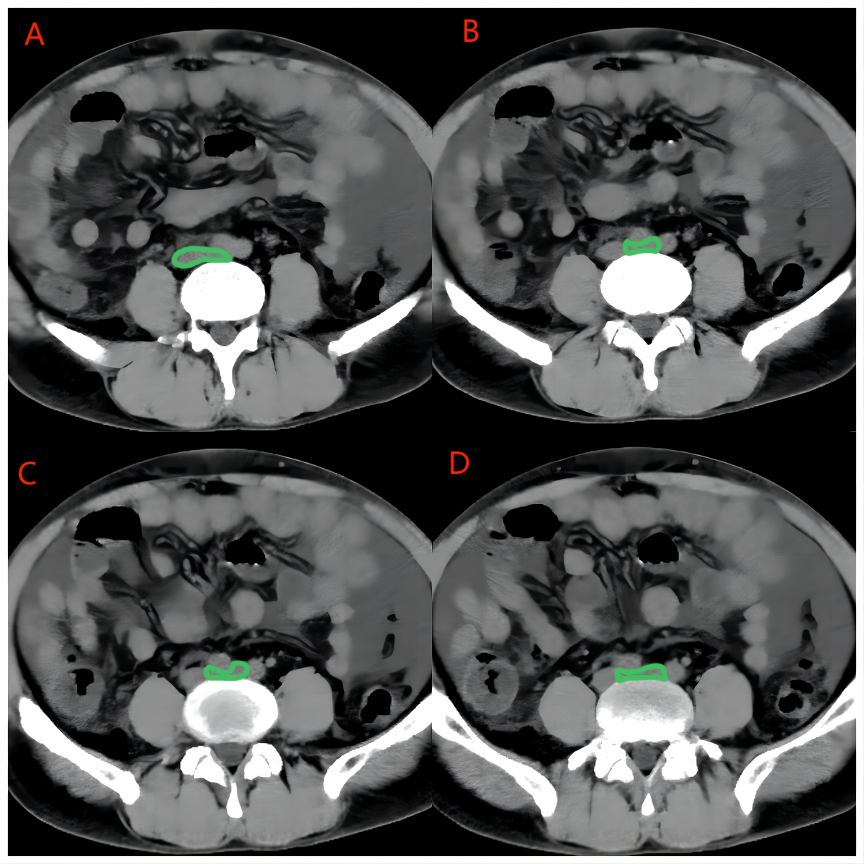}
\caption{Pic. A-B: The green mark is the bifurcation of the inferior vena cava, so that the left common iliac vein is significantly compressed by the right common iliac artery and the fifth lumbar spine; Pic. C-D: the left common iliac vein is significantly compressed, and the estimated stenosis rate is greater than 80\%.}
\label{cockett_severe}
\end{figure}

\paragraph{\hspace{1em}a) MTS-CT Dataset}  The self-collected dataset\footnote{Ethical approval (L2024018) was obtained from the hospital ethical committees.} consists of 747 subjects diagnosed with May-Thurner syndrome, mainly sourced from patients with chronic lower extremity venous insufficiency (varicose veins and edema) attending our hospital's vascular surgery department. Some patients also have coexisting Cockett syndrome. Each subject comprises 10 to 12
frames.  Notably, 366 of these subjects possess both CT scan images and their corresponding Enhanced-CT scan images. MTS-CT has a balanced gender distribution, comprising 396 males and 351 females, with an average patient age of 62.4 years, and a higher
prevalence in younger adults. All patients gave consent for CT and venous ascending phlebography, and the study was approved by the hospital's ethics committee. This is a single-center observational study. Furthermore, the MTS-CT Dataset provides essential 3D spatial information for MTS diagnosis, complementing the limitations of ultrasound and IVUS. Its large-scale contrast-enhanced CT scans enable reliable detection of iliac vein compression patterns, supporting both clinical practice and AI model development. The CT images were obtained using the United Imaging uCT 960+ system, which features a 320-slice, 640-detector array with 16 cm coverage and a spatial resolution of 30 lines/cm. It supports high-precision anatomical display, with standard reconstruction layer thickness $\leq$5 mm and multi-planar reconstruction (MPR) $\leq$1 mm. Contrast agent was administered at 80-100 mL (1.0-1.2 mL/kg body weight) with a flow rate of 2.5-4 mL/s, followed by saline. Scan timing was based on arterial and venous phases: arterial phase 30-35 seconds after injection, and venous phase 50 seconds later. The dataset is categorized into light-to-moderate (see Fig.~\ref{cockett_light}) and moderate-to-severe (see Fig.~\ref{cockett_severe}) stenosis groups based on the area stenosis rate, calculated by comparing the diameters of the narrowest and widest iliac vein segments. A stenosis rate $\leq$50\% is labeled as negative (light-to-moderate), while $>$50\% is labeled as positive (moderate-to-severe). The surgeons who participated in the data evaluation and classification were those who had been engaged in vascular surgery for more than 10 years and had extensive experience in reading and analyzing CT images. The grading standard for the degree of stenosis was to use the diameter ratio of the narrowest part to the widest part of the iliac vein on the cross-sectional CT and then square it to estimate the area stenosis rate of the iliac vein stenosis. The data was then retested and verified by another expert at a sampling ratio of 1 in 10 cases to ensure its controllability.

\begin{figure}[ht]
  \centering
  \includegraphics[width=\columnwidth]{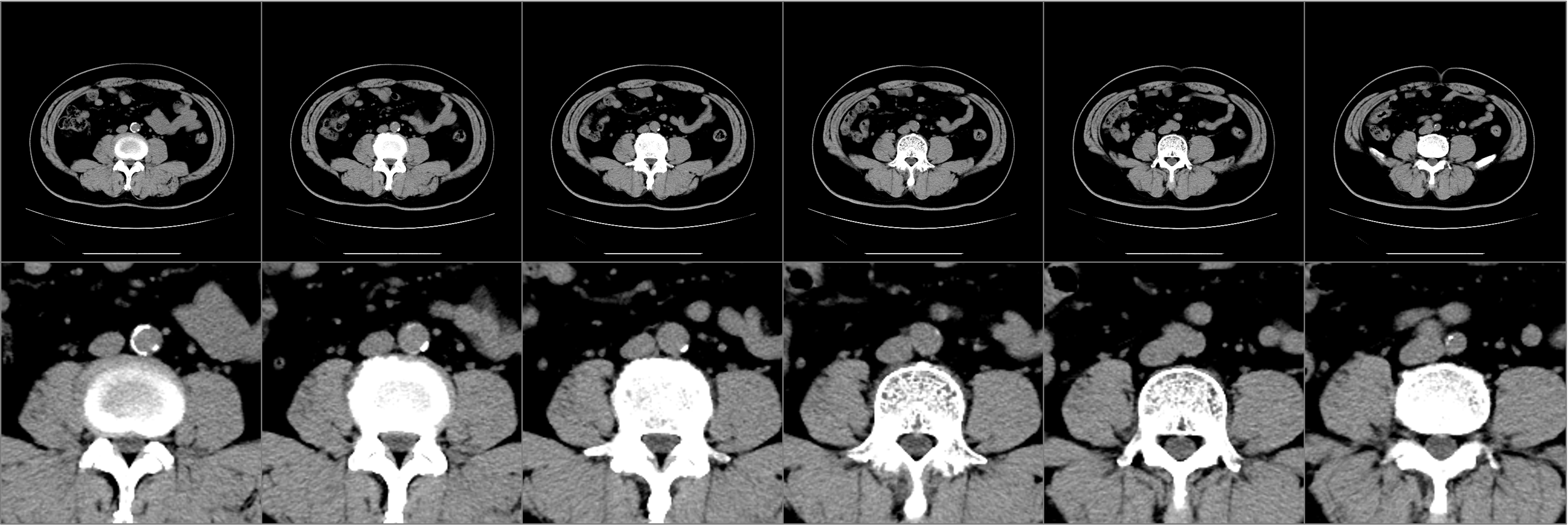}
  \caption{The first row is a subset of CT scans from a subject labelled as light-to-moderate before center cropped. The second row is the associated images after center cropped.}
  \label{center_crop_example}
\end{figure}

\noindent{\textbf{Preprocessing.}} Each CT scan image is originally saved in DICOM format at a resolution of $512 \times 512$ pixels. The patient-specific information is anonymized and de-identified to comply with regulations regarding the sharing of medical images. Each image undergoes a transformation using a grey-scale mapping function, also known as a windowing function, with a window center of 50 and a width of 200. This transformation maps hounsfield units (HU) to standardized pixel value from $0$ to $255$. Since medical experts primarily concentrate on the lumbar vertebrae, adjacent arteries, and veins in the central region of CT scans to diagnose MTS, we therefore apply a center crop to the image, as illustrated in Figure \ref{center_crop_example}.

\noindent{\textbf{Metrics.}} In this work, we use the following metrics to evaluate our proposed method. Specifically, we employ accuracy, F1-Score and area under curve (AUC) for MTS classification of 3D CT scans under various settings. We also calculate the size of parameters to discuss the trade-off between model size and performance. 

\subsection{Implementation Details}
\label{imp_details}
We have chosen 100 subjects for our test set, ensuring a balanced positive-negative ratio of 50:50. The remaining data were utilized as training and validation set for various tasks. When integrating the DEP-MHSA attention mechanism into the network, it was only applied to the final two layers, i.e., Layer 3 and Layer 4. The earlier stages remain identical to ResNet-18: Layer 1 and Layer 2 each contain two BasicBlocks; Layer 1 keeps the spatial resolution and 64 channels, whereas the first block of Layer 2 employs a $2\times2$ spatial stride (temporal stride 1) to halve the resolution while expanding the width to 128 channels. This implementation was selected by considering the balance between augmenting the network's capability for feature extraction and maintaining computational efficiency. After Layer 4, global spatio-temporal average pooling produces a 512-dimensional feature vector that is passed to a fully connected layer Linear for the final binary classification. MTS-Net was implemented in PyTorch and trained through an NVIDIA RTX 4090 24 GB GPU with a mini-batch size of 32. An initial learning rate of $5 \times 10^{-4}$ was selected because rates of $3 \times 10^{-4}$ or $8 \times 10^{-4}$ each lowered the test accuracy by roughly 3 percentage points, indicating under- and over-aggressive updates, respectively. Likewise, we retained the 512-dimensional projection before the classifier because compressing it to 256 dimensions or expanding it to 1024 dimensions degraded the average accuracy by about 5 percentage points, which demonstrates, respectively, that the smaller bottleneck diminishes discriminative capacity whereas the larger one exacerbates over-fitting. A step decay with factor g = 0.2 was applied every 25 epochs until convergence over 100 epochs. We run each configuration 10 times and report the results on the test set with the average and standard deviation of three evaluation metrics. We established the architectural frameworks of ResNet-18 with both 3D and (2 + 1)D convolutions as baselines, with distinctions summarized as follows.

\noindent{\textbf{ResNet-18 (2+1)D}}. This design decomposes 3D convolutions into separate spatial and temporal components, as detailed in Section~\ref{model_architecture}. In the stem layer, the spatial convolution uses a $1\times7\times7$ kernel, while the temporal convolutions employs a $3\times1\times1$ kernel. For the remaining network layers, kernel sizes are adjusted to $1\times3\times3$ for spatial and $3\times1\times1$ for temporal convolutions, except for those specified in the attention modules at Layer 3 and Layer 4.

\noindent{\textbf{ResNet-18 (3D)}}. The stem layer in this design employs a convolution kernel of size $3\times7\times7$ in its Conv-BN-ReLU sequence. Subsequent convolutions throughout the network consistently apply a $3\times3\times3$ kernel except for those specified in the attention modules at Layer 3 and Layer 4.

In addition to the DEP-MHSA module, we have also implemented other attention mechanisms for comparison purpose in Table~\ref{metrics_table}. Except where specifically noted, all convolution kernels in these self-attention modules follow a $1\times1\times1$ configuration. Note that the configurations regarding different attention mechanisms and 3D convolutions are stated in Section~\ref{imp_details}.

\noindent{\textbf{3D Self-Attention}}. This mechanism was implemented in a manner akin to the multi-head self-attention 
(MHSA) methods outlined in \cite{plotka2022babynet, transformer_bottleneck}. It utilizes 3D relative positional encodings integrated into the Key matrix. In its multi-head architecture, the intermediate representation's channels are segmented by the number of heads, and the final two dimensions of $\mathbb{R}^{L \times C \times H \times W}$ are flattened to generate the Query, Key, and Value matrices. 

\noindent{\textbf{(2+1)D Self-Attention}}. In line with approaches described in \cite{vivit, ruan2022mmdiffusion}, this two-step mechanism first applies MHSA on the $H \times W$ spatial dimensions and then on the temporal (frames) dimension.
  
\noindent{\textbf{Channel-wise Vanilla Attention}}. The implementation of channel-wise attention module follows the design presented in \cite{Zamir2021MPRNet}, utilizing a Conv-ReLU-Conv-$\sigma(x)$ sequence to derive attention weights. The first Conv of the sequence is a convolution with a kernel size of $1\times3\times3$ and the second convolution uses a kernel size of $3\times1\times1$. The $\sigma (x)$ is a sigmoid function which maps outputs to the range of $(0,1)$. The output undergoes an adaptive average pooling across channels before being channel-wise multiplied with the input $x$.
  
\noindent{\textbf{(2+1)D Vanilla Attention}}. The design of this module is similar to the channel-wise attention framework. Distinctively, it comprises \textbf{two} Conv-ReLU-Conv-$\sigma(x)$ sequences. The first sequence, all convolution operation utilize $1\times3\times3$ kernels for spatial emphasis on the $H \times W$ dimensions. In contrast, the second sequence employs $3\times1\times1$ kernels, focusing on temporal features. 

\subsection{Comparison with State-of-the-Art}
\paragraph{\hspace{1em}a) Baselines} In this study, we utilize a variety of models for comprehensive benchmarking, encompassing both classic and cutting-edge deep learning approaches for the classification of May-Thurner Syndrome (MTS) using 3D CT scans. Considering the dataset size, we establish our baseline using a vanilla 3D ResNet-18~\cite{r3d_1,r3d_2}. We further analyze the impact of employing a larger model or fine-tuning with a pre-trained model \cite{chen2019med3d}. Additionally, we conduct comparative experiments with models utilizing (2+1)D convolutions\cite{r2dplus1d}, as well as those equipped with attention mechanisms. Specifically, we explore various attention mechanisms for this task such as multi-head self attention\cite{plotka2022babynet}, spatial-temporal attention, and channel-wise attention \cite{hu2018squeeze_channelAttn}. We adopted 3D self-attention implementation from \cite{plotka2022babynet}, and (2+1)D self-attention implementation from \cite{ruan2022mmdiffusion}. For DenseNet-BC($k=$32, depth$=$201), we adopt the implementation from \cite{huang2017densely, hara3dcnns}. All models are trained under consistent settings.

\paragraph{\hspace{1em}b) Human Experts} We engaged five radiologists, each with approximately ten years of experience, to interpret CT images for the 100 patients. They achieved an average diagnostic accuracy of 78.4\%. The F1-Score is not reported and the AUC score is not applicable in this case.

\noindent{\textbf{Evaluation and Analysis.}} Table~\ref{metrics_table} shows quantitative comparisons between (a) proposed method: MTS-Net and (b) baseline methods: ResNet-18 (2+1)D \cite{r2dplus1d}, ResNet-18 (3D)\cite{r2dplus1d}. We discover that without any attention mechanisms, the spatial-temporal structure of ResNet-18 (2+1)D has better performance than ResNet-18 (3D). Our baseline analysis also reveals that the integration of attention modules into neural networks does not consistently lead to enhanced performance. The most optimal outcome in our study is achieved through the implementation of the proposed method. Utilizing the spatial-temporal architecture of ResNet-18, our novel dual-enhanced positional multi-head self-attention (DEP-MHSA) mechanism attains highest average score in all metrics. Additionally, training from scratch for models with larger capacity such as ResNet-50, ResNeXt-50 \cite{resnext50} and DenseNet-BC~\cite{huang2017densely} does not guarantee performance enhancement. However, fine-tuning with the pre-trained ResNet-50 \cite{chen2019med3d} can achieve better results. In summary, models with similar depth as ResNet-18 have sufficient capability for the scope and scale of our MTS-CT dataset. This suggests that models with more layers may not necessarily yield better performance in this specific context. 

\begin{table*}
  \centering
  \captionsetup{
     justification=raggedright,
     labelfont=bf,
     singlelinecheck=off
  }
  \caption{Evaluation of the proposed method on MTS-CT dataset.} 
  \label{metrics_table}
  \adjustbox{max width=\textwidth}{
  \begin{tabular}{lllllccc}
  \toprule
  \multirow{2}{*}{Network}           & \multicolumn{2}{c}{~ ~~Self-Attention~ ~~}            & \multicolumn{2}{c}{Vanilla Attention}                 & \multicolumn{3}{c}{Metrics}                                         \\ 
  \cmidrule(lr){2-2}\cmidrule(lr){3-3}\cmidrule(lr){4-4}\cmidrule(lr){5-5}\cmidrule(lr){6-6}\cmidrule(lr){7-7}\cmidrule(lr){8-8}
                                     & \multicolumn{1}{c}{(2+1)D\cite{ruan2022mmdiffusion}} & \multicolumn{1}{c}{~ 3D\cite{transformer_bottleneck}~~} & (2+1)D      & Channel  \cite{hu2018squeeze_channelAttn}             & Accuracy             & F1-Score             & AUC                   \\ 
  \midrule
  \multirow{5}{*}{ResNet-18 (2+1)D \cite{r2dplus1d}} &   \multicolumn{1}{c}{\usym{2717}}  &   \multicolumn{1}{c}{\usym{2717}}      &  \multicolumn{1}{c}{\usym{2717}}  &  \multicolumn{1}{c}{\usym{2717}}    & $0.75 \pm 0.01$          & $0.73 \pm 0.02$          & $0.78 \pm 0.02$           \\
                                     & \multicolumn{1}{c}{$\checkmark$}     &   \multicolumn{1}{c}{\usym{2717}}        &  \multicolumn{1}{c}{\usym{2717}}  &    \multicolumn{1}{c}{\usym{2717}}  & $0.73 \pm 0.01$          & $0.72 \pm 0.02$          & $0.77 \pm 0.03$           \\
                                     &   \multicolumn{1}{c}{\usym{2717}}   & \multicolumn{1}{c}{$\checkmark$}      &  \multicolumn{1}{c}{\usym{2717}}   &  \multicolumn{1}{c}{\usym{2717}}     & $0.73 \pm 0.02$          & $0.71 \pm 0.05$          & $0.78 \pm 0.03$           \\
                                     &      \multicolumn{1}{c}{\usym{2717}}    &   \multicolumn{1}{c}{\usym{2717}}     & \multicolumn{1}{c}{$\checkmark$} &    \multicolumn{1}{c}{\usym{2717}} & $0.73 \pm 0.01$          & $0.70 \pm 0.01$          & $0.75 \pm 0.03$           \\
                                     &   \multicolumn{1}{c}{\usym{2717}}    &   \multicolumn{1}{c}{\usym{2717}}    & \multicolumn{1}{c}{\usym{2717}} & \multicolumn{1}{c}{$\checkmark$} & $0.75 \pm 0.02$          & $0.73 \pm 0.03$       & $0.79 \pm 0.03$           \\ 
  \hdashline[1pt/1pt]
  \multirow{5}{*}{ResNet-18 (3D)\cite{r2dplus1d}}    &       \multicolumn{1}{c}{\usym{2717}}   &   \multicolumn{1}{c}{\usym{2717}}      &  \multicolumn{1}{c}{\usym{2717}}    &  \multicolumn{1}{c}{\usym{2717}}  & $0.73 \pm 0.01$          & $0.72 \pm 0.01$          & $0.75 \pm 0.02$           \\
                                     & \multicolumn{1}{c}{$\checkmark$}     &    \multicolumn{1}{c}{\usym{2717}}    &    \multicolumn{1}{c}{\usym{2717}}   &  \multicolumn{1}{c}{\usym{2717}}     & $0.69 \pm 0.04$          & $0.65 \pm 0.07$          & $0.70 \pm 0.04$           \\
                                     &    \multicolumn{1}{c}{\usym{2717}}      & \multicolumn{1}{c}{$\checkmark$}      &   \multicolumn{1}{c}{\usym{2717}}    &  \multicolumn{1}{c}{\usym{2717}}      & $0.74 \pm 0.03$          & $0.72 \pm 0.05$          & $0.78 \pm 0.04$           \\
                                     &     \multicolumn{1}{c}{\usym{2717}}     &   \multicolumn{1}{c}{\usym{2717}}    & \multicolumn{1}{c}{$\checkmark$} &  \multicolumn{1}{c}{\usym{2717}}     & $0.73 \pm 0.01$          & $0.70 \pm 0.01$          & $0.75 \pm 0.02$           \\
                                     &     \multicolumn{1}{c}{\usym{2717}}    &  \multicolumn{1}{c}{\usym{2717}}     &  \multicolumn{1}{c}{\usym{2717}} & \multicolumn{1}{c}{$\checkmark$} & $0.70 \pm 0.01$          & $0.67 \pm 0.04$          & $0.72 \pm 0.02$           \\ 
  \hdashline[1pt/1pt]
  ResNet-50                          &        \multicolumn{1}{c}{\usym{2717}}       &      \multicolumn{1}{c}{\usym{2717}}     &   \multicolumn{1}{c}{\usym{2717}}      &    \multicolumn{1}{c}{\usym{2717}}     & $0.74 \pm 0.02$          & $0.74 \pm 0.03$          & $0.76 \pm 0.02$           \\
  ResNet-50 (Pretrained)\cite{chen2019med3d}             &      \multicolumn{1}{c}{\usym{2717}}      &     \multicolumn{1}{c}{\usym{2717}}          &   \multicolumn{1}{c}{\usym{2717}}   & \multicolumn{1}{c}{\usym{2717}}   & $0.76 \pm 0.02$          & $0.75 \pm 0.02$          & $0.79 \pm 0.02$           \\
  ResNeXt-50 \cite{resnext50}                        &   \multicolumn{1}{c}{\usym{2717}}     &     \multicolumn{1}{c}{\usym{2717}}      &  \multicolumn{1}{c}{\usym{2717}}  &   \multicolumn{1}{c}{\usym{2717}}   & $0.72 \pm 0.02$          & $0.70 \pm 0.02$          & $0.75 \pm 0.02$           \\
  DenseNet-BC  \cite{huang2017densely, hara3dcnns}                 &   \multicolumn{1}{c}{\usym{2717}}   &   \multicolumn{1}{c}{\usym{2717}}    &  \multicolumn{1}{c}{\usym{2717}}    &   \multicolumn{1}{c}{\usym{2717}}   & $0.73 \pm 0.01$          & $0.72 \pm 0.02$          & $0.77 \pm 0.03$           \\ 
  \hdashline[1pt/1pt]
  BabyNet \cite{plotka2022babynet}      &  \multicolumn{4}{c}{}  & 0.73 $\pm$ 0.03 & 0.71 $\pm$ 0.05 & 0.76 $\pm$ 0.03  \\ 
  \hdashline[1pt/1pt]
  Human Experts      &  \multicolumn{4}{c}{}  & 0.78 $\pm$ 0.04 & - & -   \\ 
  \hdashline[1pt/1pt]
  \textbf{MTS-Net(Ours)}         & \multicolumn{4}{c}{\textbf{Attention = DEP-MHSA}}                                                      & \textbf{0.79 $\pm$ 0.01} & \textbf{0.78 $\pm$ 0.02} & \textbf{0.84 $\pm$ 0.01}  \\ 
  \bottomrule
  \end{tabular}
  }
\end{table*}

\subsection{Ablation Study}
We conduct an ablation study to understand the effectiveness of MTS-Net with respect to the DEP-MHSA module. Table \ref{ablationTable} underscores the enhanced performance achieved through the DEP-MHSA module, which employs meticulously chosen strategies to generate $Q$,$K$,$V$ matrices and augments position information using dual-enhanced positional embeddings in self-attention. The backbone network used is ResNet-18 (2+1)D with variation in its attention module and position embedding. The MHSA(3D) refers to the multi-head self-attention module mentioned in \cite{plotka2022babynet}, and the MHSA(2+1)D denotes the self-attention module with proposed method in generating $Q$, $K$ and $V$. The DEP-Embedding indicates the dual-enhanced position embedding in the attention block. Further, Table~\ref{foursettings} shows another ablation study which validates the effectiveness of various configurations for generating the Query, Key, and Value matrices, i.e., DEP-MHSA, DEP-MHSA-B, DEP-MHSA-C and DEP-MHSA-D. Our proposed DEP-MHSA consistently outperforms other configurations, underscoring the superiority of our meticulously designed structure, which mirrors the diagnostic logic of medical experts.

\begin{table}
  \centering
  \footnotesize
  \captionsetup{justification=raggedright,   labelfont=bf,  singlelinecheck=off}
  \caption{The ablation study conducted on the proposed components of the DEP-MHSA.}
  \label{ablationTable}
  \scalebox{0.66}{
  \begin{tabular}{lllllll} 
  \toprule
  \multicolumn{3}{c}{Methods}                                                   & \multicolumn{3}{c}{Metrics}                                                           & \multicolumn{1}{c}{\multirow{2}{*}{Params}}  \\ 
  \cmidrule(lr){1-1}\cmidrule(lr){2-2}\cmidrule(lr){3-3}\cmidrule(lr){4-4}\cmidrule(lr){5-5}\cmidrule(lr){6-6}
  MHSA(3D)     & \multicolumn{1}{c}{MHSA(2+1)D} & \multicolumn{1}{c}{DEP-Embedding } & \multicolumn{1}{c}{Accuracy} & \multicolumn{1}{c}{F1-Score} & \multicolumn{1}{c}{AUC} & \multicolumn{1}{c}{}                         \\ 
  \midrule
  \multicolumn{1}{c}{$\checkmark$} &   \multicolumn{1}{c}{\usym{2717}}    &  \multicolumn{1}{c}{\usym{2717}}      & $0.73 \pm 0.02$              & $0.71 \pm 0.05$              & $0.78 \pm 0.03$         &    17.41M                                  \\
  \multicolumn{1}{c}{$\checkmark$} &   \multicolumn{1}{c}{\usym{2717}}   & \multicolumn{1}{c}{$\checkmark$}                & $0.75 \pm 0.02$              & $0.75 \pm 0.02$              & $0.79 \pm 0.02$         &  17.45M                                     \\
  \multicolumn{1}{c}{\usym{2717}}    & \multicolumn{1}{c}{$\checkmark$}                    &    \multicolumn{1}{c}{\usym{2717}}    & $0.69 \pm 0.01$              & $0.65 \pm 0.05$              & $0.74 \pm 0.01$         & 31.13M                                       \\
  \multicolumn{1}{c}{\usym{2717}}    & \multicolumn{1}{c}{$\checkmark$}                    & \multicolumn{1}{c}{$\checkmark$}                 & \textbf{0.79 $\pm$ 0.01}             & \textbf{0.78 $\pm$ 0.02}              & \textbf{0.84 $\pm$ 0.01}         & 31.17M                                       \\
  \bottomrule
  \end{tabular}
  }
\end{table}

\begin{table}
  \centering
  \footnotesize
  \captionsetup{   justification=raggedright,   labelfont=bf,  singlelinecheck=off}
  \caption{Comparing four configuration of generating matrices for self-attention}
  \label{foursettings}
  \begin{tabular}{cccc} 
  \toprule
  Variants   & Accuracy        & F1-Score        & AUC              \\ 
  \midrule
  DEP-MHSA & \textbf{0.79 $\pm$ 0.01} & \textbf{0.78 $\pm$ 0.02} & \textbf{0.84 $\pm$ 0.01}  \\
  DEP-MHSA-B & $0.75 \pm 0.02$ & $0.73 \pm 0.01$ & $0.80 \pm 0.02$  \\
  DEP-MHSA-C & $0.70 \pm 0.03$ & $0.67 \pm 0.06$ & $0.74 \pm 0.04$  \\
  DEP-MHSA-D & $0.76 \pm 0.03$ & $0.75 \pm 0.04$ & $0.82 \pm 0.03$  \\
  \bottomrule
  \end{tabular}
\end{table}

\begin{table}[h]
  \captionsetup{%
     justification=raggedright,
     labelfont=bf,
     singlelinecheck=off
  }
  \centering
  \caption{The performance comparison between CT videos and Enhanced-CT videos under the same training settings.}
  \label{zq_vs_ct}
  \begin{tabular}{llll} 
  \toprule
  Data        & Accuracy        & F1-Score        & AUC              \\ 
  \midrule
  CT          & $0.69 \pm 0.03$ & $0.65 \pm 0.03$ & $0.74 \pm 0.04$  \\
  Enhanced-CT & $0.79 \pm 0.02$ & $0.77 \pm 0.03$ & $0.84 \pm 0.03$  \\
  \bottomrule
  \end{tabular}
\end{table}

\section{Discussion}
\label{discussion}
\begin{figure}[H]
  \centering
  \includegraphics[width=0.8\textwidth]{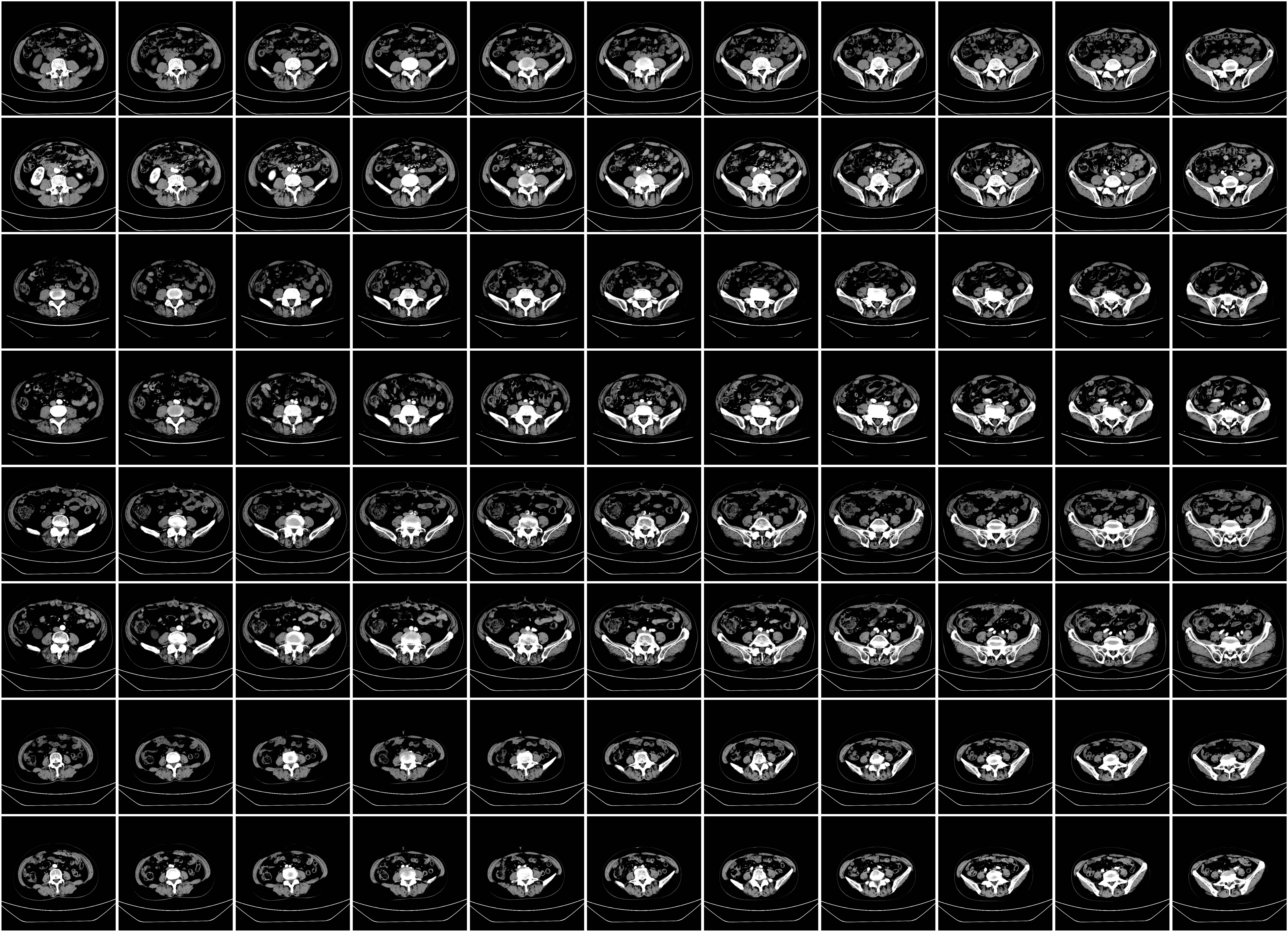}
  \caption{The figure shows a comparative display of CT and Enhanced-CT scan images with even-numbered rows depicting Enhanced-CT scans and odd-numbered rows illustrating CT scans. The Enhanced-CT images are distinctly characterized by more highlighted regions compared to the CT images. Each pair of successive rows, from top to bottom, represents CT and Enhanced-CT scans from the same subject. The first two subjects, corresponding to rows 1 through 4, are labeled as negative, while the subjects in rows 5 to 8 are labeled as positive.}
  \label{ps_vs_zq}
\end{figure}

\textbf{CT vs. Enhanced-CT.} Our study focuses on May-Thurner Syndrome classification by proposing a novel self-attention module that demonstrates superior performance using 3D CT scans. In this section, we further conduct the performance comparison between CT scans and Enhanced-CT scans through the proposed method (see Fig.~\ref{ps_vs_zq}). Enhanced-CT technique has been shown to assist medical experts in improving accuracy in diagnosing May-Thurner Syndrome compared to CT scans.  Therefore, we perform another in-depth analysis aiming to determine whether deep learning approaches can achieve similar benefits from the Enhanced-CT technique. As seen in Table \ref{zq_vs_ct}, the training set and validation set are in size of 263 in total, which is significantly less than the experiments which are exclusively carried only CT scans in Table \ref{metrics_table},\ref{ablationTable}, and Table \ref{foursettings}. With the limited training dataset, a significant performance gap is observed between the two types of data, indicating that enhanced CT scans enable better diagnostic performance with fewer training samples.

\noindent\textbf{Dataset Size.} The dataset used in this study, despite having a small number of frames for each subject, includes a relatively large number of subjects (747 subjects) compared to other popular CT scan image datasets~ \cite{dataset_ct_dataset_rister2020ct, heller2023kits19, heller2023kits21, bilic2023liver}. To our knowledge, this is the first publicly available CT scan dataset for May-Thurner Syndrome. Its significance is highlighted by the fact that around 20\% of the population are likely to have this anatomical variant \cite{medical_knowledge_phillips2023may,medical_knowledge_kalu2013may,medical_knowledge_moudgill2009may}, laying the groundwork for further studies.

\noindent{\textbf{Parameter Size.}} Table \ref{ablationTable} shows that our MHSA(2+1)D approach significantly increases the model's parameters, nearly doubling those of the standard ResNet-18(2+1)D. Performance variance is observed with and without DEP-Embedding, attributed to the disruption of relative positional information arising from distinct strategies for generating $Q$, $K$ and $V$ matrices. While these strategies enrich the model's intermediate representations, they complicate the alignment across dimensions in subsequent computations. In contrast, a uniform approach to generating $Q$, $K$ and $V$ maintains this crucial alignment. Thus, the introduction of dual-enhanced position embedding becomes necessary, making it easier to restore the lost relative positional information and enhancing performance in our MHSA(2+1)D module.

\section{Conclusions}
\label{conclusions}
We have developed MTS-Net, a novel 3D deep learning framework that significantly advances the automated diagnosis of May-Thurner Syndrome from CT scans. By integrating our specially designed Dual-Enhanced Positional Multi-Head Self-Attention (DEP-MHSA) module with a 3D ResNet-18 architecture, the system effectively captures subtle venous compression patterns through innovative multi-scale convolution and enhanced positional embeddings, closely replicating clinical diagnostic processes. We also determine the best practices of DEP-MHSA through rigorous experimental analysis on the convolution sequence in generating Query, Key and Value. To enable this research and support future studies, we've created and released the first public MTS-CT dataset, containing carefully curated scans from 747 gender-balanced subjects. Rigorous testing shows MTS-Net's strong performance with 0.79 accuracy, 0.84 AUC, and 0.78 F1-score, surpassing existing 3D CNN models including  BabyNet and DenseNet implementations by 6\%, further validating the effectiveness of the proposed DEP-MHSA. Beyond its technical innovation in vascular compression analysis, this work makes important practical contributions through its high-quality benchmark dataset and demonstration of deep learning's clinical potential for MTS diagnosis. Looking ahead, the framework shows promise for extension to other vascular compression syndromes, with future work planned for multicenter validation. We are making both the dataset and code publicly available to support continued progress in automated vascular disorder detection.

\section*{Credit Author Statement}

\textbf{Yixin Huang:} Conceptualization, Methodology, Software, Data
curation, Writing - original draft. \textbf{Yiqi Jin}: Data Curation, Validation, Investigation. \textbf{Ke Tao}: Conceptualization, Data Curation, Validation, Writing – review \& editing. \textbf{Kaijian Xia}: Validation, Writing – review \& editing. \textbf{Jianfeng Gu}: Supervision, Validation. \textbf{Lei Yu}: Validation. \textbf{Haojie Li}: Validation, Writing – review \& editing. \textbf{Lan Du}: Supervision, Validation. \textbf{Cunjian Chen}: Supervision, Validation, Writing – review \& editing.

\section*{Acknowledgment}

This research is based upon work supported in part by the Faculty Initiatives Research, Monash University, via Contract No. 2901912, in part by the Monash Suzhou Research Institute, via Contract No. MSRI8002003 and MSRI8001013, and support from the NVIDIA Academic Hardware Grant Program.

\bibliographystyle{unsrt}
\bibliography{references}

\end{document}